\documentclass[aps,twocolumn,amsmath,amssymb,preprintnumbers]{revtex4}
\usepackage{amsmath} \usepackage{amsfonts} \usepackage{amssymb}
\usepackage{bbm}
\usepackage{epsfig}
\usepackage{graphics}
\usepackage{graphicx}
\newcommand{\be}{\begin{equation}}
\newcommand{\ee}{\end{equation}}
\newcommand{\ba}{\begin{eqnarray}}
\newcommand{\ea}{\end{eqnarray}}
\newcommand{\nn}{\nonumber}

\begin{document}

\title[ ]{Growing neutrinos and cosmological selection}

\author{C. Wetterich}
\affiliation{Institut  f\"ur Theoretische Physik\\
Universit\"at Heidelberg\\
Philosophenweg 16, D-69120 Heidelberg}

\begin{abstract}
The time evolution of a cosmological scalar field can be stopped by an increasing mass of the neutrinos. This leads to a transition from a cosmological scaling solution with dynamical dark energy at early time to a cosmological constant dominated universe at late time. The trigger for the transition is set at the time when the neutrinos become non-relativistic. The characteristic mass scale for dark energy is thus related to the neutrino mass, the only known particle physics scale in its vicinity. This explains the ``why now problem'' for dark energy. We present a particle physics realization of this ``growing matter'' scenario. It is based on the very slowly varying mass of a superheavy scalar triplet field whose expectation value dominates the light neutrino masses. 
\end{abstract}

\maketitle

Growing matter has been proposed recently as a mechanism to stop the time evolution of a dynamical dark energy \cite{ABW}. An association of this growing matter component with neutrinos, ``growing neutrinos'', yields interesting relations between the present mass of the neutrinos and the dark energy density or its equation of state. A crucial ingredient is the variation of the neutrino mass with time. In this letter we present a simple particle physics model that realizes the growing neutrino scenario in a natural way. In particular, we will find a natural explanation for a large coupling between the neutrinos and the cosmological scalar field. 

Neutrino oscillations have provided convincing evidence that neutrinos are massive. However, the neutrino masses are tiny as compared to the masses of the charged fermions. This can be understood in terms of symmetries. Indeed, the standard model of electroweak interactions involves only left handed neutrinos such that no renormalizable mass term for the neutrinos is compatible with the gauge symmetry. As a further consequence of the gauge symmetry the difference between baryon number $B$ and lepton number $L$ is conserved by all renormalizable interactions. Neutrino masses can only arise from effective dimension five operators which involve two powers of the vacuum expectation value of the Higgs doublet, $d\approx 175$ GeV. They are suppressed by the inverse power of a large mass scale $M_{B-L}$ which is characteristic for $B-L$ violating effects within possible extensions beyond the standard model. 

The characteristic size of the neutrino masses, $m_{\nu,i}=H_id^2/M_{B-L}$, involves appropriate combinations of dimensionless couplings, $H_i$. Consistency with the observed oscillations requires for the mass of at least one neutrino $m_{\nu,i}\gtrsim 0.05 eV$. For $H_i$ of the order one this implies an upper bound $M_{B-L}\lesssim 6\cdot 10^{14}$  GeV \cite{CWN}. It is notable that this bound is lower than a possible scale of grand unification,  $M_{GUT}\approx 10^{16}$ GeV. This difference in scales is further enhanced if the $H_i$ are smaller than one or if the heaviest neutrino mass is larger than $0.05$ eV. We will assume here a ratio 
$M_{B-L}/M_{GUT}\gtrsim\sqrt{m_s/m_t}\approx 1/30$ as suggested by the necessity of $SU(4)_C$-breaking mass terms for the second generation quarks \cite{CWN}. This requires $H_i$ of the order one or larger, which may not seem very natural if the neutrino masses arise from an induced triplet (see below) and are in the range above $0.1$ eV. In this note we will propose a dynamical mechanism where $H$ is driven to large values in the course of the cosmological evolution.

This mechanism will lead to a fast increase of the neutrino mass from a generic value $m_\nu\lesssim 10^{-3}$ eV to its present substantially larger value. This increase is due to the time evolution of a scalar field - the cosmon - which changes its value even in the present cosmological epoch. The growing neutrino mass has rather dramatic consequences for cosmology. It essentially stops the cosmological evolution of the cosmon and triggers an accelerated expansion of the Universe, thus realizing the ``growing matter'' scenario \cite{ABW}. The increase of the neutrino mass acts as a cosmological clock or trigger for the crossover to a new cosmological epoch. 

The evolution of the cosmon field stops close to a value $\varphi_t$ which is characteristic for the transition between the two different cosmological epochs. This value does not correspond to a minimum of the effective potential $V(\varphi)$ for this scalar field. It is rather selected by a cosmological event, namely the sudden increase of the neutrino masses. The almost constant asymptotic value of the dark energy is given by $V(\varphi_t)$. It is determined by a ``principle of cosmological selection'' rather than by the properties of the vacuum. 

The most general mass matrix for the three light neutrinos reads \cite{CWN}
\be\label{1}
M_\nu=M_DM^{-1}_RM^T_D+M_L.
\ee
The first term accounts for the seesaw mechanism \cite{SS}. It involves the mass matrix for heavy ``right handed'' neutrinos, $M_R$, and the Dirac mass term $M_D=h_\nu d$. The second term accounts for the ``cascade'' or ``induced triplet'' mechanism \cite{MW}
\be\label{2}
M_L=h_L\gamma\frac{d^2}{M^2_t}.
\ee
Here a small expectation value of a heavy $SU(2)_L$-triplet field with mass $M_t$ is induced by a cubic coupling $\gamma$ involving the triplet and two powers of the Higgs-doublet. The triplet carries two units of lepton number such that $\gamma\sim M_{B-L}$. In view of the repetition of the gauge hierarchy $(d/M_{GUT})$ in the respective size of the doublet and triplet expectation values we may call this the ``cascade mechanism''. (The cascade mechanism is often called ``seesaw II'', which seems not the most appropriate name since no diagonalization of a mass matrix with large and small entries is involved, in contrast to the first term in eq. (\ref{1}).) For simplicity we will neglect here the generation structure $(M_\nu$ and $M_L$ are $3\times 3$ matrices) and associate $m_\nu$ with the average neutrino mass 
\be\label{3}
m_\nu=\frac{h^2_\nu d^2}{m_R}+\frac{h_L\gamma d^2}{M_t^2}.
\ee
With $m_R=\sigma M_{B-L}~,~h_L\gamma=\kappa M_{B-L}$ the dimensionless combination $H$, defined by $m_\nu=H d^2/M_{B-L}$, obeys $H=h^2_\nu/\sigma+\kappa M^2_{B-L}/M^2_t$. Discarding large dimensionless couplings $h_\nu,\kappa$ large values of $H$ require small $\sigma$ or small $M^2_t/M^2_{B-L}$. We will realize here the second alternative by a time dependent $M_t$, but a similar mechanism with time dependent $m_R$ is also possible. 

The generic size of the triplet mass is $M_t\approx M_{GUT}$. As a key feature of our scenario we assume that $M_t$ depends on the value of the cosmon field $\varphi$,
\be\label{4}
M^2_t=c_tM^2_{GUT}
\left[1-\frac{1}{\tau}\exp\left(-\epsilon\frac{\varphi}{M}\right)\right],
\ee
with $c_t$ and $\tau$ of the order one, $\tau>1$, and $M$ the reduced Planck mass. For $\epsilon<0$ the triplet mass decreases with increasing $\varphi$ and has a zero, $M_t(\varphi_t)=0$, for 
\be\label{5}
\frac{\varphi_t}{M}=-\frac{\ln\tau}{\epsilon}.
\ee
In consequence, the neutrino mass increases fast when $\varphi$ approaches $\varphi_t$
\be\label{6}
m_\nu(\varphi)=\bar m_\nu 
\left\{ 1-\exp\left[-\frac{\epsilon}{M}(\varphi-\varphi_t)\right]\right\}^{-1}.
\ee
Here we have replaced the parameter $\tau$ by $\varphi_t$ and we neglect the seesaw contribution, which is subleading for the range of $\varphi$ near $\varphi_t$ relevant for our discussion. The parameter $\bar m_\nu$ is given by $\bar m_\nu=(\kappa/c_t)(M_{B-L}/M^2_{GUT})d^2$, with a characteristic size $\bar m_\nu\approx 3\cdot 10^{-5}$ eV for $\kappa/c_t=1/3~,~M_{B-L}/M_{GUT}\approx 1/30$. 

For $\varphi$ near $\varphi_t$ we can approximate
\be\label{7}
m_\nu(\varphi)=\frac{\bar m_\nu M}{\epsilon(\varphi-\varphi_t)}.
\ee
Only this range will be relevant for the quantitative discussion of cosmology below. We observe that the detailed form of the $\varphi$-dependence of $M_t$ is actually not important. It is sufficient that $M^2_t(\varphi)$ crosses zero for $\varphi=\varphi_t$ and admits a Taylor expansion at this point. The neutrino mass depends only on the two effective parameters appearing in eq. (\ref{7}), namely $\varphi_t/M$ and $\bar m_\nu/\epsilon$. Since only a small range of $\varphi$ near $\varphi_t$ plays a role we can neglect the $\varphi$ dependence of all particle physics parameters except for $M_t$ or $m_\nu$. 

Before discussing cosmology we also have to specify the dynamics of the cosmon field as determined by a Lagrangian $\sim\frac12\partial^\mu\varphi\partial_\mu\varphi+V(\varphi)$. We choose an exponential cosmon potential
\be\label{8}
V(\varphi)=M^4\exp\left(-\alpha\frac{\varphi}{M}\right).
\ee
Cosmology will therefore depend on three parameters, $\alpha,\varphi_t/M$ and $\bar m_\nu/\epsilon$. In addition, the matter density at some initial time, $\rho_M(t_{eq})$, can be mapped into today's value of the Hubble parameter $H_0$. In early cosmology the neutrino mass is negligible and neutrinos behave as a relativistic fluid. Their number density is fixed, as usual, by the physics of decoupling as described by the ratio of effective neutrino and photon temperatures $T_\nu/T_\gamma$. In this early period the cosmological evolution depends only on $\alpha$ and is described by a scaling solution \cite{CWQ} with a constant small fraction of dark energy density
\be\label{9}
\Omega_{h,e}=\frac{n}{\alpha^2},
\ee
with $n=3(4)$ for the matter (radiation) dominated epoch.

However, the neutrino mass grows with increasing 
\be\label{9A}
\varphi=\varphi_0+(2M/\alpha)\ln(t/t_0).
\ee
The scaling period ends once the neutrinos become non-relativistic. Then the cosmon-neutrino coupling influences the field equation for the cosmon \cite{CWQ2}, \cite{N}
\ba\label{10}
\ddot\varphi+3H\dot\varphi&=&-\frac{\partial V}{\partial\varphi}+
\frac{\beta(\varphi)}{M}(\rho_\nu-3p_\nu),\\
\beta(\varphi)&=&-M\frac{\partial}{\partial\varphi}\ln m_\nu(\varphi)=\frac{M}{\varphi-\varphi_t}.\nn
\ea
Here $\rho_\nu$ and $p_\nu$ are the neutrino energy density and pressure, obeying
\ba\label{11}
\dot \rho_\nu+3H(\rho_\nu+p_\nu)&=&-\frac{\beta(\varphi)}{M}(\rho_\nu-3p_\nu)\dot\varphi\nn\\
&=&-\frac{\dot\varphi}{\varphi-\varphi_t}(\rho_\nu-3p_\nu).
\ea
The r.h.s of eq. (\ref{11}) accounts for the energy exchange between neutrinos and the cosmon due to the varying neutrino mass \cite{CWQ2}. We observe $\beta(\varphi)<0$ for the range $\varphi<\varphi_t$ where $\varphi$ increases towards $\varphi_t$. The effective coupling $\beta$ diverges for $\varphi\to\varphi_t$ and can therefore become very large for $\varphi$ near $\varphi_t$. This effect stops the evolution of $\varphi$ which approaches the value $\varphi_t$ arbitrarily close but cannot cross it. As a consequence, the potential energy approaches a constant, $V(\varphi)\to V_t=V(\varphi_t)$, which acts similar to a cosmological constant and causes the accelerated expansion. As $\varphi$ approaches $\varphi_t$ the kinetic energy $\dot\varphi^2/2$ must vanish asymptotically. Therefore the equation of state for the cosmon will approach the value $w_\phi=-1$, and the combined equation of state for the cosmon and neutrinos approaches $w=-V_t/(V_t+2\rho_\nu)$. We will see that $\rho_\nu/V_t$ vanishes asymptotically, such that $w\to-1$. In summary, the cosmology of our model describes a rather abrupt transition from the scaling solution for early cosmology to a cosmology with a cosmological constant 
\be\label{12}
V_t=M^4\exp\left(-\alpha\frac{\varphi_t}{M}\right).
\ee

For $\alpha\varphi_t/M\approx 276$ the cosmological constant has a value compatible with observation. This amounts to the condition (\ref{5})
\be\label{13}
\epsilon=-\frac{\alpha\ln\tau}{276}.
\ee
Upper bounds on early dark energy require $\alpha\gtrsim 10$ \cite{DRW} such that for $\ln\tau=O(1)$ only a rather mild $\varphi$-dependence of the triplet mass is required, $\epsilon \approx-0.05$. We notice that for given $\alpha$ and $\ln\tau$ a different value for $\epsilon$ would change $\alpha\varphi_t$ and therefore the present dark energy density. However, this would just change the time of the trigger event - also the neutrinos would become non-relativistic at a different time. The present dark energy density remains directly connected to the energy density of the universe at the time when the neutrinos become non-relativistic. 

From the observational bounds on the present value of the neutrino masses we infer in our scenario that this trigger event must have happened in the recent cosmological past. This allows us to relate the effective cosmological constant to the present value of the neutrino mass $m_\nu(t_0)$. For this purpose we need the present value of the cosmon field, $\varphi_0=\varphi(t_0)$. Using eqs. (\ref{7}), (\ref{10}) (with $\bar m_\nu\approx3\cdot 10^{-5}$ eV)
\be\label{14}
m_\nu(\varphi)=\frac{\beta(\varphi)}{\epsilon}\bar m_\nu
\ee
we conclude that the present value of the coupling $\beta(t_0)$ must be large and therefore $\varphi_0$ very close to $\varphi_t$. Extrapolating backwards we can compute the time $t_R$ when the neutrinos became non-relativistic. We can identify $V_t$ with the value of $V$ at the end of the early scaling solution at the ``crossing time'' $t_c$, which is close to $t_R$,
\ba\label{15}
V_t&\approx& V(t_c)\approx \frac32\Omega_{h,e}M^2H^2(t_c)\nn\\
&=&\frac{9}{2\alpha^2}M^2H^2(t_c)
=\frac{2M^2}{\alpha^2t^2_c}.
\ea
Since neutrinos have become non-relativistic only in the recent past the huge value of $t_c$ in units of the Planck time $M^{-1}$ explains the tiny present value $V(t_0)/M^4\approx 10^{-120}$. Our scenario relates the characteristic mass scale for dark energy, $\rho^{1/4}_h\approx 2\cdot 10^{-3}$ eV, to the only known particle physics scale in its vicinity, the neutrino mass. 

For a quantitative investigation of cosmological solutions it is useful to use dimensionless variables $s=-\alpha(\varphi-\varphi_t)/M~,~x=\ln a~,~\partial_t=H\partial_x$ with $V=V_t e^s$. The field equations for a homogeneous and isotropic universe can be cast into the form of evolution equations for the energy density of matter $\rho_m$, (cold dark matter and baryons), radiation $\rho_\gamma$, neutrinos $\rho_\nu$, and the cosmon $\rho_h=V+M^2\dot s^2/2\alpha^2$, 
\ba\label{16}
\partial_x\ln\rho_m&=&-3~,~\partial_x\ln\rho_\gamma=-4,\nn\\
\partial_x\ln\rho_\nu&=&-3(1+w_\nu)+\frac{\beta(s)}{\alpha}(1-3w_\nu)\partial_xs,\nn\\
\partial_x\ln\rho_h&=&-6\left(1-\frac{V}{\rho_h}\right)-\frac{\beta(s)}{\alpha}(1-3w_\nu)
\frac{\rho_\nu}{\rho_h}\partial_x s,\nn\\
\partial_xs&=&\partial_x\ln V=-\sqrt{\frac{6\alpha^2(\rho_h-V)}{\rho_h+\rho_\nu+\rho_m+\rho_\gamma}}.
\ea
Here we assume a flat universe $\rho_h+\rho_\nu+\rho_m+\rho_\gamma=3M^2H^2$, and the neutrino equation of state $w_\nu=p_\nu/\rho_\nu$ depends on the ratio between neutrino mass $m_\nu(s)$ and temperature $T_\nu$. 

The results of a numerical integration of the evolution equations are shown in figs. 1-3. As parameters we use $\alpha=10,\bar m_\nu =7\cdot 10^{-5}eV,\varphi_t/M=27.648,\epsilon=-0.05$. This results in a present neutrino mass $m_\nu(t_0)=0.44eV$ and $\Omega_m(t_0)=0.24$. Fig. 1 exhibits the crossover from a matter dominated universe to a dark energy dominated universe. We note the oscillatory behavior of the neutrino fraction $\Omega_\nu$, whereas the matter fraction $\Omega_m$ remains very smooth. The effective stop of the evolution of the cosmon field, which is triggered by the neutrinos becoming nonrelativistic, can also be seen from the equation of state in fig. 2. The fast drop in the ratio of kinetic to potential energy of the cosmon, $T/V$, (with $T=\dot\varphi^2/2$) is reflected in the fast approach of $w_h=(T-V)/(T+V)$ towards $-1$. Finally, fig. 3 shows the growing neutrino mass. Again, we observe characteristic oscillations around a smooth ``averaged value''. In the present model these oscillations seem to be stronger than for constant $\beta$ \cite{ABW}.

\begin{figure}[ht]
\begin{center}
\begin{picture}(185,155)(20,0)
\put(13,120){\small{$\Omega_m$}}
\put(210,115){$\Omega_h$}
\put(175,2){{$\ln a$}}
\put(120,25){{$\Omega_\nu$}}
\includegraphics[width=85mm,angle=0.]{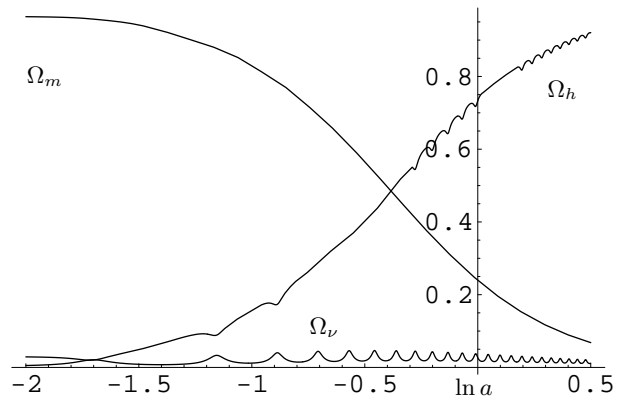}
\end{picture}
\end{center}
\caption{Fractions of matter $(\Omega_m$, highest curve on the left), dark energy $(\Omega_h=(T+V)/\rho_c$, highest curve on the right) and neutrinos $(\Omega_\nu$, small oscillating contribution), as a function of $\ln a$.}
\label{fig:figomeganl}
\end{figure}

\begin{figure}[ht]
\begin{center}
\begin{picture}(185,155)(20,0)
\put(40,130){{$z$}}
\put(150,115){{$w$}}
\put(160,30){{$w_h$}}
\includegraphics[width=85mm,angle=0.]{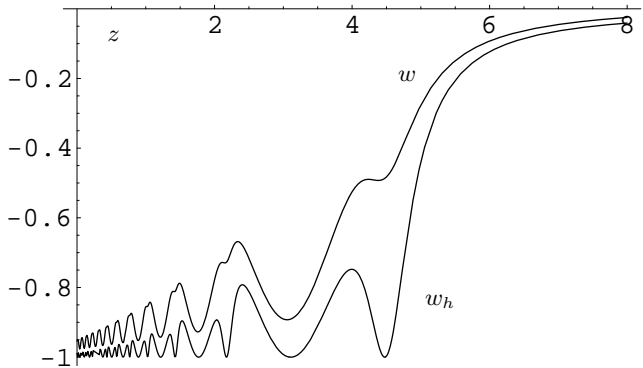}
\end{picture}
\end{center}
\caption{Equation of state of dark energy in dependence on redshift $z$. The upper curve shows the combined dark energy and neutrino sector, i.e. $w$ according to eq. (\ref{26}). The lower curve gives the pure cosmon equation of state, $w_h$, according to eq. (\ref{27}).}
\label{fig:figeosnl}
\end{figure}

\begin{figure}[t]
\begin{center}
\begin{picture}(185,155)(20,0)
\put(215,108){$\uparrow$}
\put(215,91){$\downarrow$}
\put(160,90){{$m_\nu$}}
\put(70,20){\small{}}
\put(205,2){{$\ln a$}}
\includegraphics[width=85mm,angle=0.]{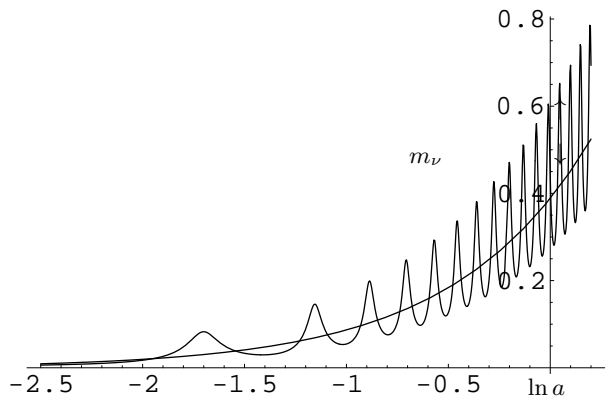}
\end{picture}
\end{center}
\caption{Growing neutrino mass a function of redshift $z$, in units of $eV$. We observe substantial oscillations around the smooth approximate solution (\ref{20}).}
\label{fig:fignymassnl}
\end{figure}

\begin{figure}[t]
\begin{center}
\begin{picture}(185,155)(20,0)
\put(150,100){$H/H_{\Lambda CDM}$}
\put(190,25){{$z$}}
\includegraphics[width=85mm,angle=0.]{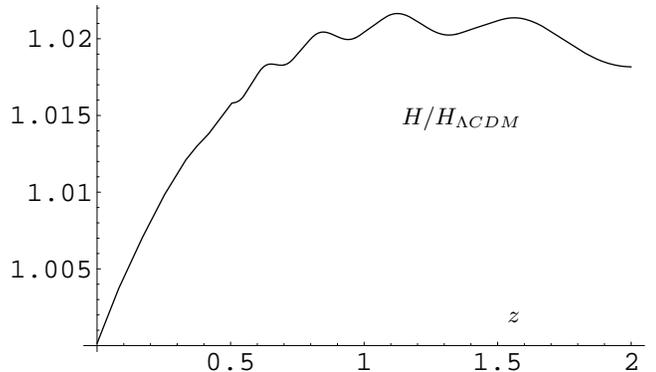}
\end{picture}
\end{center}
\caption{Hubble parameter as function of redshift. We show $H(z)$ for our model, normalized to a $\Lambda CDM$ model with the same $\Omega_m=0.24$ and massless neutrinos.}
\label{fig:fighubblenl}
\end{figure}

The late time solution can be discussed analytically. In this regime we can use eq. (\ref{10}), $\beta(s)=-\alpha/s$, and neglect the neutrino pressure, $w_\nu=0$. From
\be\label{17}
\partial_x\ln\rho_\nu+\partial_x\ln s=-3~,~\partial_x\ln\rho_m=-3
\ee
we infer
\be\label{18}
\rho_\nu=\frac{c_\nu}{sa^3}~,~\rho_m=\frac{\rho_{m,0}}{a^3}.
\ee
where $c_\nu$ has to be adjusted in order to match the early evolution. This can be inserted in the field equation (\ref{10}) for $s$ 
\be\label{19}
\partial^2_xs+\partial_x\ln H\partial_xs+3\partial_xs=-\frac{\alpha^2}{H^2M^2}
\left(V_te^s-\frac{c_\nu e^{-3x}}{s^2}\right).
\ee
The general solution of eq. (\ref{19}) describes damped oscillations around some ``average solution''. The oscillations are also visible in fig. 3.

We are interested here in the average solution for which the derivative terms on the l.h.s. of eq. (\ref{19}) are small separately. For small $s$ and $\partial_xs$ we can then proceed to an iterative solution $s(x)=s^{(0)}(x)\big(1+f(x)\big)$. For the lowest order the r.h.s. of eq. (\ref{19}) must vanish for $e^s\approx 1$, such that 
\be\label{20}
s^{(0)}(x)=\left(\frac{c_\nu}{V_t}\right)^{1/2}e^{-\frac{3x}{2}}=\frac{\tilde\rho_\nu (x)}{V_t}.
\ee
The next order indeed decays for $x\to \infty$,
\ba\label{22}
f^{(1)}(x)&=&\left(\frac{c_\nu}{V_t}\right)^{1/2}e^{-\frac{3x}{2}}
\left\{\frac{3+2\partial_x\ln H}{8\alpha^2}
\left(1+\frac{\rho_{m,0}}{V_t}e^{-3x}\right.\right.\nn\\
&&\left.\left.+2\sqrt{\frac{c_\nu}{V_t}}e^{-\frac{3x}{2}}\right)-\frac12\right\}.
\ea
At present $(x=0)$ the value of $s^{(0)}$ is already quite small
\be\label{23}
s^{(0)}_0=\left(\frac{c_\nu}{V_t}\right)^{1/2}=\frac{\tilde\rho_{\nu,0}}{V_t}\approx\frac{\Omega_\nu(t_0)}{\Omega_h(t_0)},
\ee
since the fraction of the energy density in neutrinos, $\Omega_\nu(t_0) \approx m_\nu(t_0)/16$ eV, is small. The leading order is therefore a very good approximation (note $f^{(1)}\sim s^{(0)}$). The range of validity of the smooth average solution (\ref{20}) starts actually rather early. This is demonstrated in fig. 3 where we also display the smooth average solution (\ref{20}). We observe that the actual present neutrino energy density $\rho_\nu(t_0)$ may deviate from the ``smoothened'' $\tilde\rho_{\nu,0}=\tilde\rho_\nu(t_0)$ due to oscillation effects. In the following we always will discuss smoothened quantities, often without a special distinction in the symbols.

Several cosmological observations only involve the redshift dependence of the Hubble parameter. To a good approximation its late time behavior is given by the smooth solution (\ref{20}) and reads 
\be\label{21}
H^{2}=\frac{1}{3M^2}\left\{V_t+\rho_{m,0}a^{-3}+2\tilde\rho_{\nu,0}a^{-\frac32}\right\}.
\ee
The fraction in homogeneous dark energy, $\Omega_h(t_0)\approx3/4$, consists almost entirely of potential energy $V_t$. As compared to the $\Lambda CDM$ model the only difference arises from the small fraction in neutrinos, resulting in an effective additional energy density $2\tilde\rho_{\nu,0}(a^{-3/2}-1)$. 

Correspondingly, the present equation of state for the combined growing neutrinos and the cosmon is very close to $-1$, similar to \cite{ABW}
\ba\label{26}
w=\frac{T-V+w_\nu\rho_\nu}{T+V+\rho_\nu}&\approx&-1+\frac{\rho_\nu}{V}\approx-1+\frac{\Omega_\nu}{\Omega_h},\nn\\
w_0&\approx&-1+\frac{m_\nu(t_0)}{12\rm{eV}}.
\ea
This effective dark equation of state $w$ determines the evolution of the combined cosmon and neutrino energy densities
\be\label{27A}
\partial_x\ln(\rho_h+\rho_\nu)=-3(1+w).
\ee

Incidentally, the equation of state for the cosmon alone is even closer to $-1$, as can be seen in fig. 2. For the smooth solution it would read
\be\label{27}
w_h=\frac{T-V}{T+V}~,~\tilde w_h+1\approx\frac{2\tilde T}{V}\approx\frac{16}{9\alpha^2}\Omega^2_\nu.
\ee
Indeed, we may evaluate the kinetic energy of the cosmon for the smooth leading order solution
\ba\label{24}
\tilde T&=&\frac12\dot\varphi^2=\frac{H^2M^2}{2\alpha^2}(\partial_xs^{(0)})^2\nn\\
&=&\frac{9H^2M^2c_\nu}{8\alpha^2V_t}e^{-3x}=\frac{3}{8\alpha^2}(s^{(0)})^2\rho_c.
\ea
Its present value in units of the critical energy density $\rho_c=3M^2H^2$ can be expressed in terms of the neutrino fraction $\Omega_\nu$
\be\label{25}
\frac{\tilde T}{\rho_c}=\frac{3}{8\alpha^2}
\left(\frac{\Omega_\nu}{\Omega_h}\right)^2\approx\frac{2}{3\alpha^2}\Omega^2_\nu.
\ee
However, the kinetic energy of the oscillations around the smooth solution is much higher. The average value of $T$ for low redshift exceeds by a factor of more than $10^3$ the result $\tilde T$ (\ref{27}) for the smooth solution. 

The luminosity distance probed by supernovae derivation can be directly related to the redshift dependence of $H(z)$ by integration. The latter is given by eq. (\ref{21}) for the recent epoch  and corresponds to a very mildly $z$-dependent equation of state $w(z)$. For supernovae observations or other observations of the expansion history at low redshift it seems extremely hard to distinguish our model from a cosmological constant, if the neutrino mass is small. For $m_\nu(t_0)=0.45eV$ the difference between our model and a cosmological constant (with the same $\Omega_m$ and massless neutrinos) amounts only to $1-2\%$ in the relevant redshift range, as shown in fig. 4. The difference to the $\Lambda CDM$ model increases for larger neutrino mass $m_\nu(t_0)$. 

For cosmological probes of earlier cosmology, like the CMB anisotropies or structure growth, our model comes again quite close to the $\Lambda CDM$ model if $\alpha$ is large enough. The essential difference arises from a small fraction of early dark energy, $\Omega_{h,e}\approx 3/\alpha^2$, which is absent for a cosmological constant. Such an early dark energy component influences the peak structure of the CMB-anisotropies and slows down the growth of structure. For our numerical solution we have chosen $\alpha=10$ which is close to the upper bound of a few percent for $\Omega_{h,e}$ \cite{DRW}. For larger $\alpha$ it would be even harder to distinguish our model from the $\Lambda CDM$ model. 

Due to the small size of the neutrino fraction $\Omega_\nu$ the cosmology of our model is actually very simple. To a good approximation we can simply match the early scaling solution and the late solution, neglecting the corrections from neutrinos. This means that the cosmon field increases according to the scaling solution (\ref{9A}) until it reaches $\varphi_t$. Then its evolution is suddenly frozen and $V(\varphi)$ is given by a ``cosmological constant'' $V_t=V(\varphi_t)$. The matching happens at a ``crossing time $t_c$'' or corresponding redshift $z_c$. In the matter dominated era one has
\be\label{28}
V(t)=\frac{3}{2\alpha^2}\rho_c(t)=\frac{2M^2}{\alpha^2t^2}.
\ee
In terms of the present Hubble parameter $H_0$ and dark energy fraction $\Omega_{h,0}\approx 3/4$ the condition $V(t_c)=V_t$ yields
\be\label{29}
t^2_cH^2_0=\frac{2}{3\Omega_{h,0}\alpha^2}\approx\frac{8}{9\alpha^2}.
\ee
The relation between $t_c$ and $z_c$ can be found by extrapolating the ``late cosmology'' with a cosmological constant $V_t$ backwards from the present time $t_0$ to $t_c$. For the numerical solution with $\alpha=10$ we find $z_c\approx 5$. 

The equation of state for the cosmon is well approximated by a step function
\be\label{30}
w_h(z)=-\theta(z_c-z).
\ee
At this point we notice that the precise exponential form of the potential is actually not crucial for our scenario. For a different cosmon potential one typically has $w_h(z)\approx-1$ for $z<z_c$, while the behavior of $w_h(z)$ for $z<z_c$ may differ from the scaling solution, i.e. $w_h(z>z_c)\neq 0$. Nevertheless, strong deviations from the exponential form of the potential typically need a fine tuning of parameters in order to obtain $V_t$ in the required order of magnitude. This contrasts to the exponential type which needs no such tuning by virtue of the existence of a scaling solution as a cosmic attractor. 

Neglecting $\Omega_\nu$ the cosmology depends only on two parameters, namely $\alpha$ and $\varphi_t/M$. The role of the third parameter, $\bar m_\nu/\epsilon$, concerns only the properties of the small fraction in neutrinos, $\Omega_\nu$. If we neglect the inhomogeneous fluctuations in the neutrinos we may combine the contributions of the cosmon and the neutrinos into a common effective dark energy, $\tilde\Omega_h=\Omega_h+\Omega_\nu$. Its dependence on the scale factor $a=(1+z)^{-1}$ can be approximated (after matter-radiation equality) as
\ba\label{31A}
\tilde\Omega_h(a)=\left\{\begin{array}{cl}
\frac{\tilde\Omega_{h,0}a^3+2\Omega_{\nu,0}(a^{3/2}-a^3)}{1-\tilde\Omega_{h,0}(1-a^3)+2\Omega_{\nu,0}(a^{3/2}-a^3)}&
{\rm for~}a>a_c\\
~&~\\
\frac{3}{\alpha^2}&{\rm for~}a<a_c\end{array}
\right.
\ea
where $a_c(\alpha,\tilde\Omega_{h,0},\Omega_{\nu,0})$ is determined by continuity of $\tilde\Omega_h(a)$. (For our numerical example one has $a_c=0.17$ and the approximate formulae (\ref{31A}) underestimates $\tilde\Omega_h$ by at most $25\%$ in a small region of $a$ around $a_c$.) Eq. (\ref{31A}) can be used to estimate luminosity distances and other quantities of the background cosmology in terms of four parameters: the present value of the Hubble parameter $H_0$ or $h$, the present fraction in dark energy $\tilde\Omega_{h,0}$, the present value of the neutrino fraction, which is related to the average neutrino mass by $\Omega_{\nu,0}=m_\nu(t_0)/(30.8 h^2$ eV), and the early dark energy fraction $\Omega_{h,e}=3/\alpha^2$. Here we recall that the laboratory value of the neutrino mass $m_\nu(t_0)$ may differ from the average value $\tilde m_\nu(t_0)$ due to the oscillations around the smooth  solution. For a given neutrino mass the only parameter in addition to the $\Lambda CDM$-model is $\alpha$. The effective dark energy equation of state obtains then by taking a derivative 
\ba\label{31B}
w&=&\frac13 \partial_x\ln(\tilde\Omega^{-1}_h-1)\\
&=&\left[-1+\frac{\Omega_{\nu,0}a^{-3/2}}{\tilde\Omega_{h,0}+2\Omega_{\nu,0}(a^{-3/2}-1)}\right]
\theta(a-a_c)\nn
\ea
and we recover eq. (\ref{30}) for $\Omega_{\nu,0}\to 0$.

We observe that the fraction in neutrinos remains small during the whole cosmological history since matter domination. It decreases while the neutrinos are relativistic, increases once the neutrinos become non-relativistic due to their increasing mass, and finally decreases again in the period dominated by the effective cosmological constant. Despite the small value of $\Omega_\nu$ the neutrinos are important for the evolution of the cosmon due to their large coupling $\beta(\varphi)$. As compared to neutrinos with constant mass the time $t_R$ when the neutrinos become non-relativistic is much later in our model. With $T_\nu=T_{\nu,0}/a~,~T_{\nu,0}=1.7\cdot 10^{-4}$ eV, 
$m_\nu=\tilde m_{\nu}(t_0)a^{3/2},m_\nu(a_R)=3T_\nu(a_R)$ one finds $a_R$ somewhat smaller than in \cite{ABW}
\be\label{34}
a_R=\left(\frac{\tilde m_\nu(t_0)}{3T_{\nu,0}}\right)^{-\frac25}=0.05
\left(\frac{\tilde m_\nu(t_0)}{eV}\right)^{-2/5}.
\ee
For our numerical example $(m_\nu(t_0)=0.44 eV)$ we define $z_R$ by the condition $w_\nu(z_R)=0.15$ and find $z_R=10$. 

The growth of neutrino fluctuations for $a>a_R$ has similar qualitative properties as discussed in \cite{ABW}. Due to the extended relativistic period the neutrino fluctuations behave much closer to massless neutrinos as compared to a situation with constant $m_\nu$. Once a laboratory value $m_\nu(t_0)$ will be measured, it could well turn out to be above the cosmological bounds inferred for constant mass neutrinos. This could constitute an important test for our scenario. Possible inhomogeneous neutrino fluctuations on large scales could be a distinctive feature of our model and are under investigation. For the time being our model seems consistent with all cosmological observations.

Let us now turn back to the particle physics aspects of our scenario. First we note that the dependence of the present neutrino mass on the parameter $\bar m_\nu/\epsilon$ is not linear. For the late time solution the approximate vanishing of the r.h.s. of the first eq. (\ref{10}) can be used to express $\beta(a)$ in terms of the neutrino number density $n_\nu(a)~,~\rho_\nu-3p_\nu=m_\nu n_\nu$, i.e.
\be\label{36A}
\beta(a)m_\nu(a)n_\nu(a)\approx M\frac{\partial V}{\partial\varphi}\approx-\alpha V_t.
\ee
With eqs. (\ref{7}), (\ref{10})
\be\label{36B}
m_\nu(a)=\beta(a)\frac{\bar m_\nu}{\epsilon}
\ee
one finds
\be\label{36C}
\beta(a)=-(\alpha V_t)^{1/2}
\left(-\frac{\bar m_\nu}{\epsilon}\right)^{-1/2}n_\nu(a)^{-1/2}.
\ee
This yields the dependence of $m_\nu(a)$ on the parameter $\bar m_\nu/\epsilon$ and the known neutrino number density
\be\label{36D}
m_\nu(a)=(\alpha V_t)^{1/2}\left(-\frac{\bar m_\nu}{\epsilon}\right)^{1/2}
n_\nu(a)^{-1/2}.
\ee

It is interesting to estimate the factor by which the neutrino mass has increased since early times
\ba\label{31}
\frac{\tilde m_\nu(t_0)}{\bar m_\nu}&=&\frac{M}{\epsilon(\varphi_0-\varphi_t)}=-\frac{\alpha}{\epsilon s(t_0)}\nn\\
&=&-\frac{\alpha\Omega_h(t_0)}{\epsilon\Omega_\nu(t_0)}=\frac{\beta(\varphi_0)}{\epsilon}.
\ea
For $\alpha=10,\epsilon=-0.05~,~\Omega_{h,0}=\frac34$ this yields $\tilde m_\nu(t_0)/\bar m_\nu\approx 150/\Omega_\nu(t_0)$.  This can be used for an estimate of $\bar m_\nu$ as a function of the present neutrino mass $\tilde m_\nu(t_0)$ (assuming $h=0.72)$
\ba\label{32}
\bar m_\nu&=&-\frac{\epsilon}{\alpha\Omega_h(t_0)}
\frac{\tilde m^2_\nu(t_0)}{16eV}\nn\\
&\approx&\frac{1}{2400}
\left(\frac{\tilde m_\nu(t_0)}{eV}\right)^2eV,
\ea
yielding $\bar m_\nu=(10^{-7}, 4\cdot 10^{-5},2\cdot 10^{-3})$ eV for $\tilde m_\nu(t_0)=(0.015,0.3,2.3)$ eV. This fits well with the particle physics estimates at the beginning of this letter. One may turn this argument around and use the particle physics estimate of $\bar m_\nu$ for a prediction of the present cosmological value of the neutrino mass
\be\label{35A}
\tilde m_\nu(t_0)\approx 0.3eV.
\ee
The uncertainties in the estimate of $\bar m_\nu$ are substantial, however, even though their effect on the value of $\tilde m_\nu(t_0)$ is reduced by the square root dependence in eq. (\ref{36D}). Furthermore, the laboratory value $m_\nu(t_0)$ differs from $\tilde m_\nu(t_0)$ due to the oscillatory behavior. While oscillations are moderate for $\tilde m_\nu(t_0)\approx 0.5 eV$ they become strong and very peaked for small neutrino masses, e.g. $m_\nu(t_0)\approx 0.015 eV$. 

In summary, the combination of a weakly $\varphi$-dependent triplet mass (\ref{4}) (or a corresponding $\varphi$-dependent right handed neutrino mass $m_R(\varphi)$) with a cosmon potential of the exponential type (\ref{8}) yields a rather interesting cosmology which seems consistent with the present status of observations and particle physics expectations.

In this context, we should point out that in presence of several scalar fields the cosmon potential $V(\varphi)$ corresponds to the ``valley'' where all other scalar fields have been ``integrated out'' by solving their field equations as functions of $\varphi$. In our case we have a general effective potential $U(\varphi,d,t)$ which depends also on the doublet and triplet fields $d$ and $t$ and is supposed to include all quantum fluctuations. Solving the $\varphi$-dependent field equations for the doublet and triplet, $\partial U/\partial d=0,~,\partial U/\partial t=0$, specifies the valley given by $d(\varphi)~,~t(\varphi)$. The cosmon potential obtains then as  $V(\varphi)=U(\varphi~,~d(\varphi)~,~t(\varphi)$). An exponential decay of $V(\varphi)$ along the valley has been motivated by considerations of the dilatation anomaly and fixed point behavior \cite{CWQ,9}. Such a fixed point is reflected in an apparent ``tuning'' of individual contributions to $V(\varphi)$, as characteristic for deviations from a fixed point for a generic renormalization flow. 

This can be demonstrated by looking at a simple potential
\ba\label{35}
U&=&U_0(\varphi)+\frac\lambda2(d^2-d^2_0)^2+\frac12 M^2_t(\varphi)t^2-\gamma d^2 t,\nn\\
M^2_t(\varphi)&=&\bar M^2_t\left[1-\exp \left(-\frac\epsilon M(\varphi-\varphi_t\right)\right].
\ea
Inserting the solution for the triplet $t=\gamma d^2/M^2_t(\varphi)$ yields
\be\label{36}
U\big(\varphi,d,t(d,\varphi)\big)=U_0(\varphi)+\frac\lambda2(d^2-d^2_0)^2-\frac{\gamma^2d^4}{2M^2_t(\varphi)}.
\ee
The resulting $\varphi$-dependence of $d$
\be\label{37}
d^2(\varphi)=d^2_0\left(1-\frac{\gamma^2}{\lambda M^2_t(\varphi)}\right)^{-1}
\ee
yields
\be\label{38}
V(\varphi)=U_0(\varphi)-
\frac{\lambda d^4_0\gamma^2}{2(\lambda M^2_t(\varphi)-\gamma^2)}.
\ee
As long as $M^2_t\gg \gamma^2/\lambda$ the variation of $d$ is small, $d\approx d_0$, and we can approximate
\be\label{41}
V(\varphi)=U_0(\varphi)-\frac{m_\nu(\varphi)d^2\gamma}{2h_L}.
\ee
The second term is of the order $d^4$ and much larger than the present value of $V$, demonstrating the apparent ``tuning'' characteristic for fixed points. Similar considerations also hold for the derivatives of $\varphi$, including the cosmon mass term $m^2_\phi=\partial^2V/\partial\varphi^2$. Of course, the fixed point behavior is a crucial assumption of our scenario that needs to be established in a suitable unified theory.

The $\varphi$-dependence of $d$ results in a time variation of the quark and charged lepton masses and the Fermi scale of weak interactions (in units of a fixed Planck mass $M$). For an appropriate time independent Yukawa coupling we may estimate the size of this effect for the mass of the electron
\ba\label{42}
\partial_t\ln m_e&=&\frac12\partial_t\ln d^2=\frac12 \dot\varphi\partial_\varphi\ln d^2\nn\\
&=&-\left(1-\frac{\gamma^2}{\lambda M^2_t}\right)^{-1}
\frac{\gamma^2}{2\lambda M^2_t}\dot\varphi\partial_\varphi\ln M^2_t.
\ea
Concentrating on the recent cosmology of $\varphi$ near $\varphi_t$ we may compare
\be\label{43}
M^2_t =-\frac{\epsilon c_t s}{\alpha}M^2_{GUT}=-\frac{\epsilon c_t }{\alpha}\frac{\rho_\nu}{V_t}M^2_{GUT}
\ee
with the estimates after eq. (\ref{6})
\be\label{44}
\frac{\gamma^2}{\lambda}\approx10^{-4}\frac{c^2_t}{h^2_L\lambda}M^2_{GUT}.
\ee
With $(-\epsilon/\alpha)\approx 5\cdot 10^{-3}$ the ratio $R=\gamma^2/(\lambda M^2_t)$ remains small at present if $c_t/h^2_L\lambda$ is sufficiently small. One infers
\be\label{45}
\partial_t \ln m_e\approx-\frac R2\partial_t\ln s\approx -\frac R2\partial_t\ln \rho_\nu
\approx\frac{3R}{4}H.
\ee
Compatibility with present bounds requires a small value of $R$. The increase of the variation $\partial\ln m_e/\partial\ln a=3R/4$ with $R$ is due to the divergence of $M^{-2}_t$ for $\varphi \to\varphi_t$ and therefore characteristic for triplet-induced variations.

Without going to a detailed discussion this demonstrates that time varying fundamental constants are an interesting issue in the scenario where the cascade mechanism dominates the present neutrino mass. For variations induced by the triplet variation - as $d(\varphi)$ in our case - we find an increase of the variation per $e$-folding towards the present time since $R$ increases. This contrasts with variations of other couplings that are not directly related to $M_t(\varphi)$. Indeed, for generic variations one rather expects a slowing down of the variation in the recent epoch since the evolution of te cosmon almost stops. Growing neutrinos dominated by the cascade mechanism therefore suggest a comparatively strong variation of all fundamental constants related to the Fermi scale. We note that the variation effects may be smaller if we realize our scenario with a varying mass of the heavy right handed neutrinos, since the variation of $m_R(\varphi)$ does not necessarily strongly affect the couplings of the standard model.

For couplings not related to $M_t(\varphi)$ we find indeed a decrease of the variation per $e$-folding for late cosmology $(z<z_c)$. Quite generally, if a quantity $X$ depends exponentially on $\varphi/M$ we expect a slowing down of its time evolution as $\varphi$ approaches $\varphi_t$. According to 
\ba\label{46}
\frac{\partial\ln X}{\partial \ln a}&=&\frac{\partial\ln X}{\partial\varphi}\frac{\partial\varphi}{\partial\ln a}=
\frac{\delta}{M}
\frac{\partial\varphi}{\partial\ln a}=-\frac{\delta}{\alpha}\frac{\partial s}{\partial\ln a}=
\frac{3\delta}{2\alpha}s\nn\\
&=&\frac{3\delta}{2\alpha}\frac{\Omega_\nu}{\Omega_h}\approx
\frac{m_\nu(t_0)}{8{\rm eV}}\frac{\delta}{\alpha}a^{-3/2}
\ea
this yields for late cosmology a variation at redshift $z$ (for constant $\delta/\alpha$)
\be\label{47}
\frac{\delta X}{X}=-\frac{m_\nu(t_0)}{12{\rm eV}}
\frac{\delta}{\alpha}
\big((1+z)^{3/2}-1\big).
\ee
This may be compared with early cosmology $(z>z_c)$ where the scaling solution implies $(n=4(3)$ for radiation (matter) domination)
\be\label{48}
\frac{\partial\ln X}{\partial\ln a}=\sqrt{3n}\frac{\delta}{\alpha}~,~
\frac{\delta X}{X}=-\sqrt{3n}
\frac{\delta}{\alpha}\ln (1+z).
\ee
From the successful description of nucleosynthesis we draw the lesson that a realistic small time variation of fundamental constant requires a suppressed $\varphi$-dependence for the couplings in the observable sector (small $\delta/\alpha)$, perhaps to be explained by a fixed point behavior \cite{9} or by different effective mass scales $M/\alpha$ and $M/\delta$ appearing in the $\varphi$-dependence of the potential and in the observable sector. 

Let us conclude with a short discussion of the ``principle of cosmological selection''. In particle physics, the values of fundamental parameters like the Fermi scale etc. are fixed by the properties of the vacuum or ground state. In our scenario this does not hold for the value of the effective cosmological constant $V_t$. The cosmon potential $V(\varphi)$ has no minimum for $\varphi=\varphi_t$ - in the ground state the value of $\varphi$ is infinite, with $V(\varphi\to\infty)=0$. Nevertheless, a cosmological mechanism stops the evolution of the scalar field at $\varphi_t$. This is due to the growing mass of excitations - in our case the neutrinos. Cosmological selection is conceptually quite different from the ground state selection in particle physics. It would be interesting to see if cosmological selection could also play a role for the determination of some other parameter in the standard model of particle physics.


\newpage
\begin{thebibliography}{100}
\bibitem{ABW}L. Amendola, M. Baldi, C. Wetterich, arXiv: 0706.3064 [astro-ph]
\bibitem{CWN}C. Wetterich, Nucl. Phys. {\bf B187} (1981) 343
\bibitem{SS}P. Minkowski, Phys. Lett. {\bf B67} (1977) 421;\\
T.~Yanagida, in: Workshop on Unified Theory, (KEK report 79-18, 1979), eds O.~Savada and A.~Sugamato, p95;\\
M.~Gell-Mann, P.~Ramond, R.~Slansky, in: Supergravity, eds. D.~Freedman and P.~van Niewenhuizen, North Holland, 1979
\bibitem{MW}M. Magg, C. Wetterich, Phys. Lett. {\bf 94B} (1980) 61\\
G. Lazarides, Q. Shafi, C. Wetterich, Nucl. Phys. {\bf B181} (1981) 287; 
R.~N.~Mohapatra, G.~Senjanovic, Phys. Rev. {\bf D23} (1981) 165;\\
J.~Schechter, J.~W.~F.~Valle, Phys. Rev. {\bf D22} (1980) 2227
\bibitem{CWQ}C. Wetterich, Nucl. Phys. {\bf B302} (1988) 668
\bibitem{CWQ2}C. Wetterich, Nucl. Phys. {\bf B302} (1988) 645
\bibitem{N}P. Gu, X. Wang, X. Zhang, Phys. Rev. {\bf D68}, 087301 (2003);\\
R.~Fardon, A.~E.~Nelson, N.~Weiner, JCAP0410 (2004), 005; A.~W.~Brookfield et al., Phys. Rev. Lett. {\bf 96} (2006) 061301; N.~Afshordi, M.~Zaldarriaga, K.~Kohri, Phys. Rev. {\bf D72} (2005) 065024; O.~E.~Bjaelde et al., arXiv: 0705.2018; K.~Ichiki, Y.~Keum, arxiv: 0705.2134
\bibitem{DRW}M. Doran, G. Robbers, C. Wetterich, Phys. Rev. {\bf D75} (2007) 023003
\bibitem{9}C. Wetterich, Phys. Rev. Lett. {\bf 90} (2003) 231302; Phys. Lett. {\bf B561} (2003) 10
\end{thebibliography}
\end{document}